# Darker than black: radiation-absorbing metamaterial

E. E. Narimanov

*Birck Nanotechnology Center, Purdue University, West Lafayette, IN 47907*

H. Li, Yu. A. Barnakov, T. U. Tumkur, M. A. Noginov*

*Center for Materials Research, Norfolk State University, Norfolk, VA 23504*

*\* mnoginov@nsu.edu*

**Abstract:**

We show that corrugated surfaces of hyperbolic metamaterials scatter light preferentially inside the media, resulting in a very low reflectance and ultimate dark appearance in the spectral range of hyperbolic dispersion. This phenomenon of fundamental importance, demonstrated experimentally in arrays of silver nanowires grown in alumina membranes, originates from a broad-band singularity in the density of photonic states. It paves the road to a variety of applications ranging from the stealth technology to high-efficiency solar cells and photodetectors.

*Metamaterials* – engineered composite materials with rationally designed subwavelength inclusions – have unique responses to electromagnetic fields, which are unavailable in conventional media. Their fascinating properties and applications include negative index of refraction [1,2], invisibility cloaking [3-5], and sub-diffraction imaging and focusing [2,6-9]. The research in *hyperbolic metamaterials* (also known as *indefinite media* [10]), originally stimulated by tantalizing possibilities offered by the absence of diffraction limit in a hyperlens [6-9], has uncovered a number of novel effects resulting from a broadband singular behavior of the density of photonic states in these materials [11,12]. Here we show that the latter phenomenon leads to dramatic enhancement of light scattering from defects and surface corrugations, with nearly all incident photons "sucked" into propagating modes of a hyperbolic medium (silver filled alumina membrane) [14].

Known hyperbolic metamaterials include arrays of metallic nanowires grown in dielectric membranes [13-15] as well as lamellar metal-dielectric or semiconductor structures [16-18]. These uniaxial materials are highly anisotropic, with the dielectric permittivity components of opposite signs in two perpendicular directions. An isofrequency dispersion curve for an extraordinary wave in a uniaxial (meta)material is given by

$$\frac{k_y^2}{\varepsilon_{x,z}} + \frac{k_x^2 + k_z^2}{\varepsilon_y} = \left(\frac{\omega}{c}\right)^2, \qquad (1)$$

where $k_i$ are the components of the wave-vector ($i=x,y,z$; the optical axis is in the $y$ direction), $\varepsilon_i$ are the values of electric permittivity, $\omega$ is the angular frequency, and $c$ is the speed of light.

The nature of the "super-singularity" in hyperbolic metamaterials can be understood from a visual representation of the density of states in terms of the phase space volume enclosed by two surfaces corresponding to different values of the light frequency. In dielectric media, electric permittivities are positive, $\varepsilon_i>0$, and the dispersion law describes an ellipsoid in the k-space (sphere if $\varepsilon_x=\varepsilon_y=\varepsilon_z$). The phase space volume enclosed between two such surfaces (Fig. 1(a), left) is then finite, corresponding to a finite density of photonic states. However, when one of the components of the dielectric permittivity tensor is negative, Eq. (1) describes a hyperboloid in the phase space. As a result, the phase space volume between two such hyperboloids (corresponding to different values of frequency) is infinite (Fig. 1(a), right), leading to an infinite density of photonic states. While there are many mechanisms leading to a singularity in the density of photonic states, this one is unique – as it results in the infinite value of the density of states for every frequency where different components of the dielectric permittivity have opposite signs.

As the scattering rate $W_{k \to k'}$, in the spirit of the Fermi Golden Rule [19], is proportional to the density of states $\rho(\omega)$,

$$W_{k \to k'} \propto \delta(\omega_k - \omega_{k'})\rho(\omega),  \qquad (2)$$

the super-singularity in $\rho(\omega)$ dramatically enhances light scattering from surface defects into the bulk of a hyperbolic metamaterial, with the resulting suppression of the light reflection. We therefore predict that, contrary to naive intuitive expectations, introducing defects (such as *e.g.* surface corrugations) at the surface of a hyperbolic metamaterial should dramatically reduce the total light reflection, including the diffuse component [20].

Hyperbolic metamaterials, which we used in our experiments, were based on arrays of 35 nm-thick silver nanowires grown (using an electroplating technique) in 1cm x 1cm x 51μm anodic alumina membranes [15]. Nanowires were oriented perpendicularly to the membrane's surface, and the nominal filling factor of silver was ~12%. The fabricated uniaxial metamaterial reported here had hyperbolic dispersion (negative electric permittivity in the direction perpendicular to the membrane's surface, $\varepsilon_\perp$, and positive electric permittivity in the direction parallel to the membrane, $\varepsilon_\parallel$) in the near-infrared spectral range. From the measurements of the angular dependence of reflectance in s and p polarizations, Fig. 1(b), we have deduced the following materials parameters at $\lambda$=873 nm: $\varepsilon_\perp'$=-0.15, $\varepsilon_\perp''$=1.07, $\varepsilon_\parallel'$=4.99, and $\varepsilon_\parallel''$=0.02. The measurements and the retrieval procedure are discussed in detail in Ref. [15]. The roughness of the untreated sample, measured with the atomic force microscope (AFM), was equal to rms=40 nm, Fig. 1(c). The lateral correlation length, at which the correlation function

$$C(r_m,r_n) = \frac{\sum_m \sum_n (z_m(r_m)-\bar{z})(z_n(r_n)-\bar{z})}{\sum_m (z_m(r_m)-\bar{z})^2}$$ decreased from 1 to 1/e, was equal to ~100 nm, Fig. 2. (Here $z_i$ is the height of the topology profile in the position $r_i$ and $\bar{z}$ is the mean height).

The surface of the sample was then corrugated by grinding it with 3 μm $Al_2O_3$ polishing powder, resulting in the roughness equal to rms=600 nm and the correlation length equal to 1.17 μm, Fig. 1(d). The reflectance measurements were repeated in the corrugated sample, resulting in much smaller intensities of reflected light, especially at small incidence angles, Fig. 1(b). Although the reduction of reflectance was observed in both polarizations, it was stronger in p polarization than in s polarization, inset of Fig. 1b. The fact that the reduced reflection was observed not only for p-polarized but also for s-polarized incident light suggests change of the state of polarization upon scattering, which is commonly observed in scattering media [21]. The reduced sample's reflectance correlates with its enhanced transmittance measured in a spectrophotometer setup, Fig. 3.

As corrugated samples produced not only specular reflection but also diffused scattering, we have evaluated the ratio of reflected and scattered light intensities by measuring optical signal in the direction of reflected beam at different distances from the sample, inset of Fig. 4. The dependence of the detected light intensity on the distance $l$ between the sample and the large-aperture detector (integrating sphere with 1 inch diameter opening) is shown in Fig. 4. The diffused scattering, characterized by an increase of the signal at small distances $l$, has been observed in the corrugated sample and was practically absent in the untreated one. By fitting the experimental result to a simple model, it has been shown that the ratio of scattering and specular light intensities is much smaller than the difference between the measured reflectances in the corrugated and not corrugated samples. Thus, according to the most conservative estimate (assuming Lambertian scattering distribution diagram), the ratio of diffusely scattered light intensity to specularly reflected light intensity in the corrugated sample did not exceed 3.5. This value is much smaller than the ratio of reflectances in the untreated and the corrugated samples, ~30 at small angles, see Fig. 1b and inset.

We thus have demonstrated that the reflectance of a hyperbolic metamaterial is significantly reduced upon corrugation of its surface, in an agreement with the theoretical prediction. With the original concept equally applicable to all parts of the electromagnetic spectrum, our result thus opens an entirely new route towards radiation-absorbing materials and surfaces, with wide range of applications spanning from solar light harvesting to radar stealth technology.

**Acknowledgements:** The authors cordially thank D. Dryden, G. Natara for the assistance with experiments. H.L., Y.A.B., T.T., and M.A.N. acknowledge the support by the NSF PREM grant # DMR 0611430, NSF NCN grant # EEC-0228390, NSF IGERT grant #DGE 0966188, and AFOSR grant # FA9550-09-1-0456.

Figure 1.

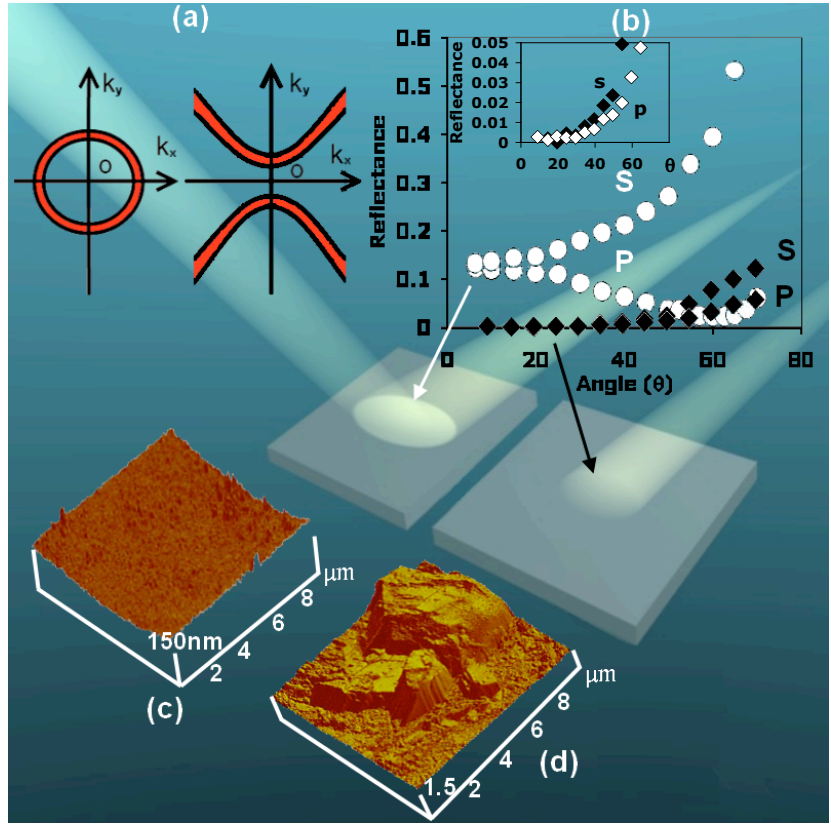

Fig. 1. DRAMATIC REDUCTION OF REFLECTANCE OFF CORRUGATED HYPERBOLIC METAMATERIAL. Panel (a): the phase space "volume" enclosed by two different surfaces of constant frequency, in the cases when components of the dielectric permittivity tensor are all positive (left) and have opposite signs (right). Panel (b): angular reflectance profiles measured on untreated (circles) and roughened (diamonds) parts of the same membrane sample in s-polarization and p-polarization. Inset: reflectance profiles in the corrugated sample (same as in main panel (b), zoomed). Panels (c) and (d): topography profiles of the untreated (c) and corrugated (d) samples.

Figure 2.

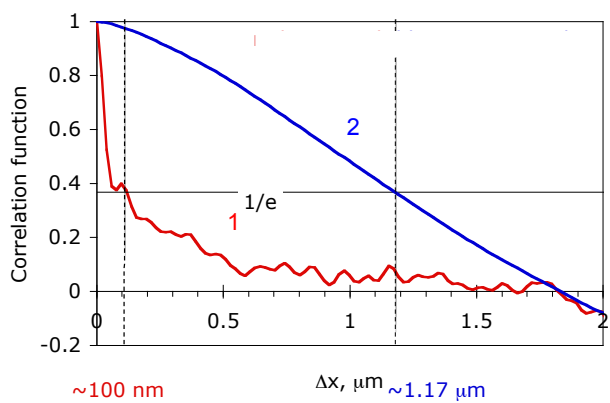

Fig. 2. Correlation functions characterizing surface roughness of the untreated (1) and corrugated (2) samples.

Figure 3.

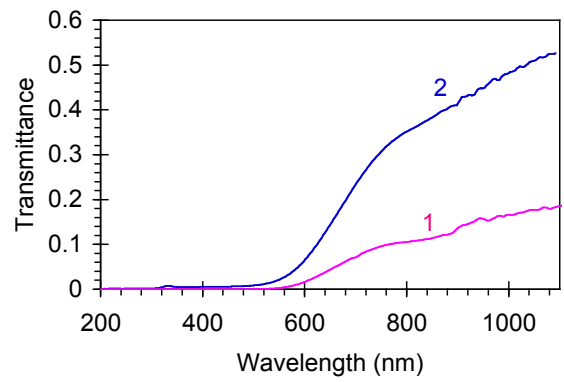

Fig. 3. Transmittance spectra of the untreated (trace 1) and corrugated (trace 2) samples measured at near-normal incidence.

Figure 4.

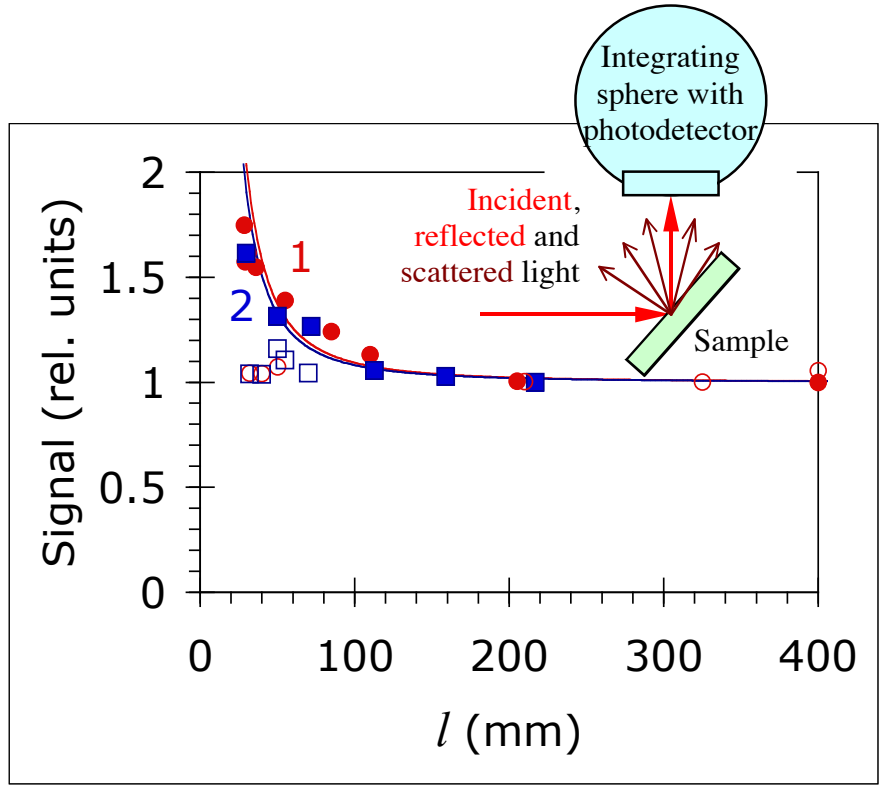

Fig. 4. Reflected and scattered light intensities measured as a function of the sample-to-detector distance $l$ with a wide-aperture detector in the untreated (open characters) and corrugated (closed characters) samples; circles – s-polarization, squares – p-polarization. The measurements were done at the 41° reflectance angle. Red and blue solid lines are the fittings of the experimental data (in corrugated samples) with the Lambertian model; trace 1 – s polarization, trace 2 – p polarization. Inset: Schematic of the experimental setup.